\documentstyle[epsf]{elsart}
\journal{Physics Letters B}
\newcommand{\mrm}[1]{\mbox{\rm #1}}
\newcommand{\bla}{\hspace{1cm}}
\newcommand{\rfn}[1]{(\ref{#1})}
\newcommand{\db}{\hspace{-0.2ex}\not\hspace{-0.7ex}D\hspace{0.1ex}}
\newcommand{\sla}[1]{\hspace{-0.1ex}\not\hspace{-0.5ex} #1\hspace{0.1ex}}
\newcommand{\mafigura}[4]{
  \begin{figure}[hbtp]
    \begin{center}
      \epsfxsize=#1 \leavevmode \epsffile{#2}
    \end{center}
    \caption{#3}
    \label{#4}
  \end{figure} }

\newcommand{\beq}{\begin{equation}}
\newcommand{\eeq}{\end{equation}}
\newcommand{\bea}{\begin{eqnarray}}
\newcommand{\eea}{\end{eqnarray}}
\newcommand{\bes}{\begin{eqnarray*}}
\newcommand{\ees}{\end{eqnarray*}}   
\newcommand{\eq}[1]{eq.~(\ref{#1})}

\newcommand{\nn}{\nonumber}
\newcommand{\Eq}[1]{Eq.~(\ref{#1})}
\newcommand{\ea}{{\it et al.}}

\newcommand{\np}[1]{Nucl. Phys. {\bf #1}}
\newcommand{\pl}[1]{Phys. Lett. {\bf #1}}
\newcommand{\pr}[1]{Phys. Rev. {\bf #1}}
\newcommand{\prl}[1]{Phys. Rev. Lett. {\bf #1}}
\newcommand{\zp}[1]{Z. Phys. {\bf #1}}
\newcommand{\prep}[1]{Phys. Rep. {\bf #1}}
\newcommand{\rmp}[1]{Rev. Mod. Phys. {\bf #1}}

\def\lsim{\mathrel{\vcenter{\hbox{$<$}\nointerlineskip\hbox{$\sim$}}}}

\def\meg{$\mu \rightarrow e\ \gamma$}
\def\mec{$\mu$--$e$ conversion}  
\def\mte{$\mu \rightarrow e e e$}

\begin{document}

\begin{frontmatter}

\hfill FTUV/97-56

\hfill IFIC/97-58

\title{$\mu$--$e$ conversion in nuclei versus $\mu \rightarrow e\ \gamma$: \\
an effective field theory point of view}
\author{Martti Raidal\thanksref{martti}} and 
\author{Arcadi Santamaria\thanksref{arcadi}}
\address{Departament de F\'{\i}sica Te\`orica, IFIC, 
CSIC-Universitat de Val\`encia,\\ E-46100 Burjassot, Val\'encia, Spain} 
\thanks[martti]{E-mail: raidal@titan.ific.uv.es}
\thanks[arcadi]{E-mail: Arcadi.Santamaria@uv.es}

\begin{abstract}
Using an effective lagrangian description we analyze possible new
physics contributions to the most relevant muon number violating
processes: $\mu \rightarrow e\ \gamma$  and
$\mu$--$e$ conversion in nuclei. We identify a general class of models 
in which those processes are generated at one loop level and in which 
$\mu$--$e$ conversion is enhanced with respect to $\mu \rightarrow e\ \gamma$ 
by a large  $\ln(m^2_\mu/\Lambda^2),$ where $\Lambda$ is the scale 
responsible for the new physics. For this wide class 
of models bounds on $\mu$--$e$ conversion constrain the scale of 
new physics more stringently than
$\mu \rightarrow e\ \gamma$ already {\it at present} and, with the
expected improvements in $\mu$--$e$ conversion experiments,
will push it upwards by about one order of magnitude more.
To illustrate this general result we give an explicit model containing
a doubly charged scalar and derive new bounds on its couplings to 
the leptons.
\end{abstract}
\end{frontmatter}

\section{Introduction}

The precision reached in the last years in the experiments searching for
\mec{}  in nuclei at PSI \cite{psiti,psipb} and 
TRIUMF \cite{triumph} and the expected improvement in the sensitivity 
of the experiments at PSI in the next years 
by more than two  orders of magnitude \cite{imprti} will make 
\mec{} the main test of muon flavour  conservation for
most of the extensions of the standard model (SM). 
Moreover, according to the recent BNL proposal \cite{proposal}
further improvements in the experimental sensitivity down to the level
$10^{-16}$ are feasible. 

There are three processes for 
which there are very good 
experimental bounds: \meg{}, \mec{} and \mte{}. None of them occurs in the 
SM without extending it with right-handed neutrinos or extra scalars 
\cite{reviews} but in general
they appear in any extension of the SM in which lepton flavour
conservation is not imposed by hand. \meg{}, having the photon on mass-shell, 
can only be originated from an off-diagonal (in generation space) 
magnetic moment.
This interaction can only appear, in any
renormalizable model, at the loop level. On the contrary, \mec{} and \mte{} 
can be generated at three level from renormalizable interactions by exchange of
scalars or gauge bosons. However, in many models those couplings conserve
muon flavour or do not appear in both lepton and quark sectors. In this case
\mec{} and/or \mte{} are also generated at the loop level.  If this is the
case, and given the present experimental accuracy, one often finds 
\cite{vergados} that 
the bounds on new physics coming from \meg{} 
are stronger than the bounds found from \mec{} or \mte{}. 

However, this does not need to be always
the case. It could happen that the form factors contributing
to \mec{} are enhanced with respect to the ones contributing to \meg{}. 
Indeed, it has been noticed already in the early works 
of ref.~\cite{altarelli} 
that in some cases \mec{} constrains new physics more stringently than \meg{}. 
In this Letter we investigate the conditions in which this happens in the 
framework of the effective quantum field theory  which allows one to classify 
in a simple way the contributions coming from a large variety of  extensions
of the SM. We point out a general class of models in which \mec{} 
is enhanced by
large logarithms. As an example we will study a simple
extension of the SM with a doubly charged scalar singlet 
coupled to the right-handed leptons and derive new limits on its couplings.
The same bounds, although derived for a singlet,  hold with a good precision 
also for models containing scalar triplets which are introduced usually to
generate Majorana neutrino masses \cite{triplets} and appear naturally
in left-right models \cite{lr}.
Finally we comment on other presently 
popular models (e.g., with broken $R$-parity \cite{rparity} 
or leptoquarks \cite{leptoquarks}) with the similar feature.

\section{Effective lagrangian description of theories}

Assuming that the relevant physics responsible for muon-number
nonconservation occurs at some scale
$\Lambda > \Lambda_{F} \equiv$~Fermi~scale, we can write the relevant
effective lagrangian at low energies as
\def\lla{{\cal L}^L}
\def\lra{{\cal L}^R}
\def\lsigmal{{\cal L}^{\sigma L}}
\def\lsigmar{{\cal L}^{\sigma R}}
\def\lsigma{{\cal L}^{\sigma}}
\def\lll{{\cal L}^{LL}}
\def\lrr{{\cal L}^{RR}}
\begin{equation}
\label{eq:leff}
{\cal L}_{eff} = \lla +  \lra + \lsigmal + \lsigmar + \lll + \lrr + \cdots, 
\end{equation}
where
\begin{equation}
\label{eq:la}
\lla =
\frac{\alpha^{L}_{ij}}{(4\pi)^2 \Lambda^2} e\ \overline{e_{iL}} 
\gamma_\nu e_{jL} \partial_\mu F^{\mu\nu}\,,
\end{equation}

\begin{equation}
\label{eq:ra}
\lra=
\frac{\alpha^{R}_{ij}}{(4\pi)^2 \Lambda^2} e\ \overline{e_{iR}} 
\gamma_\nu e_{jR} \partial_\mu F^{\mu\nu}\,,
\end{equation}

\begin{equation}
\label{eq:sigmal}
\lsigmal=
\frac{\alpha^{\sigma L}_{ij}}{(4\pi)^2 \Lambda^2} e\ \overline{e_{iL}} 
\sigma_{\mu\nu} i \db e_{jL} F^{\mu\nu} + \mathrm{h.c.}\,,
\end{equation}

\begin{equation}
\label{eq:sigmar}
\lsigmar=
\frac{\alpha^{\sigma R}_{ij}}{(4\pi)^2 \Lambda^2} e\ \overline{e_{iR}} 
\sigma_{\mu\nu} i \db e_{jR} F^{\mu\nu} + \mathrm{h.c.}\,,
\end{equation}


\begin{equation}
\label{eq:ll}
\lll=
\frac{\alpha^{LL}_{ik;lj}}{\Lambda^2} 
(\overline{e_{iL}} e_{kL}^c)(\overline{e_{lL}^c} e_{jL})\, ,
\end{equation}

\begin{equation}
\label{eq:rr}
\lrr=
\frac{\alpha^{RR}_{ik;lj}}{\Lambda^2} 
(\overline{e_{iR}} e_{kR}^c)(\overline{e_{lR}^c} e_{jR})\, .
\end{equation}











Here, as in the rest of the paper, we will
assume that repeated indices, Lorentz, $\mu,\nu,$  or generation indices,
$i,j,k,l$, are summed. When possible we will use also matrix notation
in generation space. $e_{iL}$ and $e_{iR}$ are chiral charged-lepton fields,
$e_{L,R}^c = (e_{L,R})^c$ are the charge conjugated fields and
$\db = \sla{\partial} + i e \sla{A}$.~The four-fermion couplings
$\alpha^{LL}_{ik;lj}$ and $\alpha^{RR}_{ik;lj}$ are symmetric with
respect to the exchanges $i \leftrightarrow k$ and/or $l \leftrightarrow j$. 

We expect that the terms $\lla$, $\lra$, $\lsigmal$, $\lsigmar$ are generated
at one loop in the renormalizable theories since they cannot be obtained from
renormalizable vertices at tree level, that is the reason we already
included a factor $(4\pi)^2$ in the denominator. $\lla$ and $\lsigmal$
will arise, for instance, in models with an extra scalar triplet with
hypercharge 1 \cite{triplets,megtriplet} or an extra
scalar singlet \cite{zee,misha} coupled to the leptonic doublet.
$\lra$ and $\lsigmar$ will arise, for instance, in models with a
doubly charged scalar singlet \cite{doublycharged} coupled to the
singlet right-handed leptons, we will study this model more carefully
latter on. 

Note the particular form we have written the magnetic-moment
type operators involving only chiral fields. The two operators 
$\lsigmal$ and  $\lsigmar$ could be combined by using the equations of
motion for the lepton fields. We have
\begin{equation}
i \db e_L = M_e e_R\ , \ \
i \db e_R = M^\dagger_e e_L\ ,
\label{eq:motion}
\end{equation}
where $M_e$ is the charged lepton mass matrix and we have used a matrix 
notation to suppress generation indices.
By using the equations of motion, which is perfectly allowed in an 
effective lagrangian \cite{eqmotion} at the lowest order, we obtain 
\begin{equation}
\lsigma \equiv \lsigmal+\lsigmar = 
\frac{1}{(4\pi)^2 \Lambda^2} e\ 
\overline{e_L} \sigma_{\mu\nu} F^{\mu\nu} \left(\alpha^{\sigma L} M_e+
M_e \alpha^{\sigma R}\right) e_R  + \mathrm{h.c.} 
\label{eq:magmoment}
\end{equation}
In fact for applications we will use $\lsigma$ written in this form.
Note that we could use from the beginning this form for $\lsigma$ as 
the starting effective lagrangian, but this is not completely 
equivalent to what we did. By doing that we would have no 
reason to choose the particular form of magnetic moments proportional
to the fermion masses present in \eq{eq:magmoment}. However, in 
chiral theories, like the ones we want to consider, magnetic moments appear
always in the form \eq{eq:magmoment} and are proportional to the fermion 
masses. In more general theories with
chirality explicitly broken independently of the fermion masses, operators
like \eq{eq:magmoment} but with an arbitrary matrix could arise. 

Note also the form in which we have written the four fermion operators
in terms of conjugate fields. Those operators are equivalent,
after a Fierz transformation, to the usual vector-vector four fermion
operators, for instance
\begin{equation}
(\overline{e_{iR}} e_{kR}^c)(\overline{e_{lR}^c} e_{jR})=
\half (\overline{e_{iR}} \gamma_\mu e_{jR})
(\overline{e_{kR}}\gamma^\mu e_{lR})\, .
\label{eq:fierz}
\end{equation}
We have chosen this form because it is simpler for loop calculations since
it leads to only one penguin diagram while the vector-vector interaction
leads to two types of penguin diagrams. Moreover, these operators
arise naturally in the class of models we will consider.

Four-charged-fermion interactions violating generation-number conservation 
however can be generated easily at tree level
in a large class of models (excluding supersymmetry with conserved R-parity).
For instance $\lll$ will appear in scalar models in which the scalars
couple to the lepton doublet and models with a scalar triplet
with hypercharge 1. Notice that a singly charged singlet cannot generate
these kind of couplings, it only generates couplings with two charged leptons
and two neutrinos \cite{zee,misha}. $\lrr$ will appear in models with a doubly 
charged singlet (more on this later). Four-fermion couplings involving both, 
charged leptons and neutrinos have not been included because, as we will
see later, they do not lead 
to any logarithmic enhancement of the \mec{} rate. Four-fermion couplings
involving leptons and quarks could also be included, if not bounded already
by another reasons, and would lead to a similar logarithmic enhancement
to the one we are going to study.

In addition to the couplings we have considered there could be
four fermion operators involving both left-handed and right-handed fields.
They could be generated, for instance,  by exchange of Higgs doublets.
However, these couplings are usually suppressed by the masses of the
fermions. The consequences of these kind of operators at tree level have
already been considered in \cite{ng1}.
Therefore, for simplicity, we are not going to consider them in this paper.
Moreover, a direct $Z_\mu  \overline{e_i} \gamma^\mu e_j$  vertex can
in principle be generated at tree level in some models in which ordinary
fermions mix with other fermions with exotic hypercharges and at one loop
in models with non-decoupling physics in the same way that extra couplings
$Z_\mu  \overline{b_L} \gamma^\mu b_L$ arise in the SM when one tries to
make the top-quark mass very heavy \cite{zbbar}. In particular those 
couplings will arise when a Dirac mass term for the neutrinos is made 
large~\cite{nondec}.
In these kind of models \mec{} could be sizable with respect to
\meg{} studied in ref. \cite{pont}. 
Since this mechanism has already been studied elsewhere 
\cite{nondec,zprim}
we are not going to consider it here anymore and will concentrate in 
models in which \mec{} proceeds through the photonic mechanism, that is,
by exchange of a photon between the leptonic and the hadronic currents.

It is important to note that the lagrangian \rfn{eq:leff} has to
be interpreted as a lagrangian in the effective field theory 
approach\footnote{For an detailed explanation of this approach in
the case of a model with a singly charged scalar see for
instance \cite{misha}.}. This means that four-fermion interactions, which
are generated at tree level, can be used at one loop and will 
generate non-analytical contributions to the electromagnetic
form factors. In fact, as we will see, those non-analytical 
contributions are quite independent of the model and are the key
of the possible enhancement of the form factors contributing to 
\mec{} with respect to those contributing to \meg{}. If there are
logarithmic contributions to the \mec{} rate, they will dominate, and 
since they can be computed in the effective theory, they are quite
independent on the details of the full theory from which the effective 
lagrangian is originated. In section \ref{sec:model} we show, by using an 
explicit model, how this works and how the \mec{} rate is quite independent 
on the details of the model.

\section{$\mu$--$e$ conversion  versus $\mu\rightarrow e\gamma$}

Theory of \mec{} in nuclei was first studied by Weinberg and Feinberg 
in ref. \cite{wf}. Since then
various nuclear models and approximations are used in the literature to 
calculate coherent \mec{} nuclear form factors. It is important to note that 
the results from the shell model \cite{ver1}, local density approximation
\cite{chiang} as well as the quasi-particle RPA approximation \cite{ver2}
do not differ significantly from each other for both 
$_{22}^{48}Ti$ and $_{82}^{208}Pb$
nuclei showing consistency in the understanding of the nuclear physics 
involved \cite{ver2}. We follow the notation of ref.~\cite{chiang} and take
into account corrections to the local density approximation from the exact 
calculations performed in the same work.
The corrections are negligible for $_{22}^{48}Ti$ as the local density 
approximation works better for light nuclei but are sizable for 
$_{82}^{208}Pb.$

The relevant \mec{} matrix element can be expressed as
\beq
{\cal M} \,=\,\frac{4\pi\alpha}{q^2} \, j^{\mu}_{(1)} J^{(1)}_{\mu}
\, + \, \frac{G_F}{\sqrt{2}} \, j^{\mu}_{(2)} J^{(2)}_{\mu}\,,
\label{me}
\eeq
where  $q$ is the momentum transfer and in a good approximation 
$q^2\approx -m^2_\mu.$ The first term in \eq{me} describes  photonic
and the second term  non-photonic conversion mechanisms.
$j^{\lambda}_{(1,2)}$ and $J^{\lambda}_{(1,2)}$ represent 
the leptonic and hadronic currents, respectively. 
The non-photonic mechanism is mediated by  heavy particles and,
therefore, suppressed compared with the photonic mechanism. 
The non-photonic mode is of interest if the conversion can occur at 
tree level like in models with non-diagonal $Z'$ \cite{zprim} and
Higgs  \cite{ng1} couplings, models with broken $R$-parity \cite{kim},
models with leptoquarks \cite{davidson} or  if the
loop contributions are enhanced by some other mechanism, e.g., models
with non-decoupling of massive neutrinos \cite{nondec}. Since we do
not consider tree level lepton number violation via four fermion operators
involving quarks in this
Letter we shall concentrate in the following on the photonic mechanism only.

Generally, the leptonic current for the photonic mechanism can be
pa\-ra\-me\-tri\-zed as 
\bea
j^{\lambda}_{(1)} &=& {\bar u}(p_e) \left[ \; 
\left (f_{E0} + \gamma_5 f_{M0}\right) \gamma_{\nu}  \left (
g^{\lambda \nu} - \frac {q^{\lambda} q^{\nu}}{q^2}\right )
\right. \nn \\
&  & +
\left. 
(f_{M1} + \gamma_5 f_{E1})\; i\; \sigma^{\lambda \nu} \frac{q_{\nu}}{m_{\mu}} 
\right] u(p_{\mu})\,,
\label{j1}
\eea
where $p_e,\,\,p_\mu$ are the lepton momenta and 
the form factors $f_{E0}$, $f_{E1}$, $f_{M0}$ and 
$f_{M1}$ can be computed from the underlying theory.
The coherent \mec{} branching ratio $R_{\mu e}$ can be expressed as \cite{chiang}
\beq
R_{\mu e}=C\,\frac{8\pi\alpha^2}{q^4}\,p_e E_e \,
\frac{|F(p_e)|^2}{\Gamma_{capt}}\,\xi_0^2\, ,
\eeq
where $E_e$ is the electron energy, $\Gamma_{capt}$ is the total
muon capture rate, $|F(p_e)|^2$ is the nuclear matrix element squared
and 
\beq
\xi_0^2=|f_{E0}+f_{M1}|^2+|f_{E1}+f_{M0}|^2
\label{eq:xi}
\eeq
shows the \mec{} dependence on the form factors. 
The expression for  $|F(p_e)|^2$ in the local density approximation,
the correction factors $C$ to the approximation (compared with the 
exact calculation) as well as  all numerical values of the above
defined quantities for  $^{48}_{22}Ti$ and $^{208}_{82}Pb$  can be found in
refs.~\cite{chiang,chiang2}. The result reads
\beq
R_{\mu e}=C\,\frac{8\alpha^5\,m_\mu^5\,Z^4_{eff}\,Z\,|\overline{F_p}(p_e)|^2}
{\Gamma_{capt}}
\cdot \frac{\xi_0^2}{q^4}\, ,
\label{mecrate}
\eeq
where $C^{Ti}=1.0,$ $C^{Pb}=1.4,$ $Z^{Ti}_{eff}=17.61,$ $Z^{Pb}_{eff}=33.81$,
$\Gamma_{capt}^{Ti}=2.59\cdot 10^6$~s$^{-1}$,
$\Gamma_{capt}^{Pb}=1.3\cdot 10^7$~s$^{-1}$ and the proton nuclear 
form factors are $\overline{F_p}^{Ti}(q)=0.55$ and
$\overline{F_p}^{Pb}(q)=0.25.$
Presently SINDRUM II experiment is running on gold \cite{imprau}, 
$_{79}^{179}Au$,
 but gold is not explicitly treated in ref. \cite{chiang}.
However, since $Z^{Au}=79$ and $Z^{Pb}=82$ are so close to each other
then, within errors, all the needed quantities for   $_{79}^{179}Au$
and $^{208}_{82}Pb$ are approximately equal\footnote{We thank 
H.C. Chiang and E. Oset for clarifying us this point.}. 
This result is strongly supported by theoretical calculations and 
experimental measurements of the total muon capture rate of $Pb$ and $Au$
\cite{chiang2}. In the following we use the same experimental and 
theoretical input for both    $_{79}^{179}Au$ and $^{208}_{82}Pb.$

One should note that the \meg{} branching ratio,
\beq
R_\gamma=\frac{96\pi^3 \alpha}{G_F^2m_\mu^4}
\left( |f_{M1}|^2+|f_{E1}|^2 \right)\, ,
\label{megrate}
\eeq
depends on a different combination of the form factors.

\mafigura{8cm}{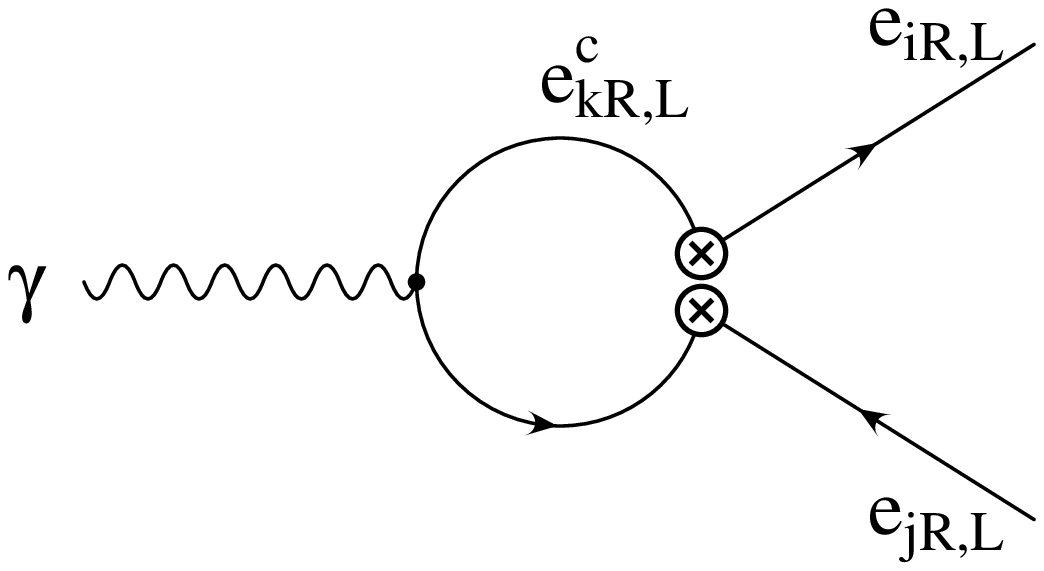}{The one-loop contribution to the electromagnetic
current in the effective theory.}{fig:effective-loop}

To show the power of the effective lagrangian description of new physics
we compute the form factors starting from the lagrangian \rfn{eq:leff}.
The one loop diagram giving rise to \mec{} is depicted
in  fig.~\ref{fig:effective-loop}.
By using the $\overline{\mathrm{MS}}$ renormalization scheme and choosing
the renormalization scale, $\mu=\Lambda$, where $\Lambda$ is 
the scale of new physics, we find for the form factors:
\bea
f_{E0}=&\frac{-q^2}{(4\pi)^2 2\Lambda^2}\bigg[
\frac{4}{3}\left(\alpha^{LL}_{ke;k\mu}+\,\alpha^{RR}_{ke;k\mu}\right)
&\bigg(\ln\frac{-q^2}{\Lambda^2}+F(r_k)\bigg)\nonumber \\
&&+\left(\alpha^{L}_{e\mu}+\,\alpha^{R}_{e\mu}\right)\bigg]\,, \label{eq:fv}\\
f_{M0}=&\frac{-q^2}{(4\pi)^2 2\Lambda^2}\bigg[
\frac{4}{3}\left(\alpha^{LL}_{ke;k\mu}-\,\alpha^{RR}_{ke;k\mu}\right)
&\bigg(\ln\frac{-q^2}{\Lambda^2}+F(r_k)\bigg)\nonumber \\
&&+\left(\alpha^{L}_{e\mu}-\,\alpha^{R}_{e\mu}\right)\bigg] \,,
\label{eq:fa}\\
f_{M1}=&\frac{m^2_\mu}{(4\pi)^2\Lambda^2}
\left(\alpha^{\sigma L}_{e\mu}+\,\alpha^{\sigma R}_{e\mu}\right)\,,&
\label{eq:sigmav}\\
f_{E1}=&\frac{m^2_\mu}{(4\pi)^2\Lambda^2}
\left(\alpha^{\sigma L}_{e\mu}-\,\alpha^{\sigma R}_{e\mu}\right)\,,&
\label{eq:sigmaa}
\eea
where $r_k=m^2_k/(-q^2)$, with $m_k$ being the masses of the fermions running in
the loop, and
\beq
F(r)=\ln r+4 r -\frac{5}{3}+
(1-2 r)\sqrt{1+4r}\, \ln\left(\frac{\sqrt{1+4r}+1}{\sqrt{1+4r}-1}\right)\, .
\label{f1}
\eeq 
There are three important limiting cases. 
If $r\to 0$ (i.e., $k=e$) then $F(r)=-5/3,$ 
if $r\approx 1$ (i.e., $k=\mu$) then $F(r)\approx 0.18$ 
and if $r\gg 1$ (i.e., $k=\tau$) then $F(r)=\ln r.$
The loop diagram in fig.~\ref{fig:effective-loop} 
give contributions only to the form factors $f_{E0}$ and $f_{M0}$
but not to  $f_{E1}$ and $f_{M1}.$
Because of the UV divergence find in the loop calculation
those contributions always contain a term which is proportional to 
$\ln(q^2/\Lambda^2)$ or $\ln(m_\tau^2/\Lambda^2).$
This term which is
completely {\it independent} of the details of the model that 
originate the four-fermion interaction gives a large 
enhancement for the form factors $f_{E0}$ and $f_{M0}$ while the 
enhancement is absent in the form factors $f_{E1}$ and $f_{M1}$.
Consequently, the \mec{}\ is enhanced while \meg{}\ is not. 
In this class of models, in which \mec{} is dominated by this
large logarithmic term one can neglect all the non-logarithmic
contributions which are the ones that depend on the details
of the complete theory.

\section{An explicit model}
\label{sec:model}

Let us consider for a moment an extension of the
SM by adding just a doubly charged  scalar singlet $\kappa^{++}$.
Its coupling to right-handed leptons are described by  
\begin{equation}
{\cal L}_\kappa = h_{ij} \overline{e_{iR}^c} e_{jR}\,\kappa^{++} + 
\mathrm{h.c.}
\label{eq:lk}
\end{equation}
Here the Yukawa coupling matrix $h_{ij}$ is symmetric in the generation
indices $i,j$.
From this interaction we obtain easily, see fig.~\ref{fig:tree-match},
the four-fermion interaction
\begin{equation}
\frac{1}{m_\kappa^2} h^*_{ki} h_{lj} 
(\overline{e_{iR}} e_{kR}^c)(\overline{e_{lR}^c} e_{jR})\,.
\label{eq:lk4}
\end{equation}
\mafigura{12cm}{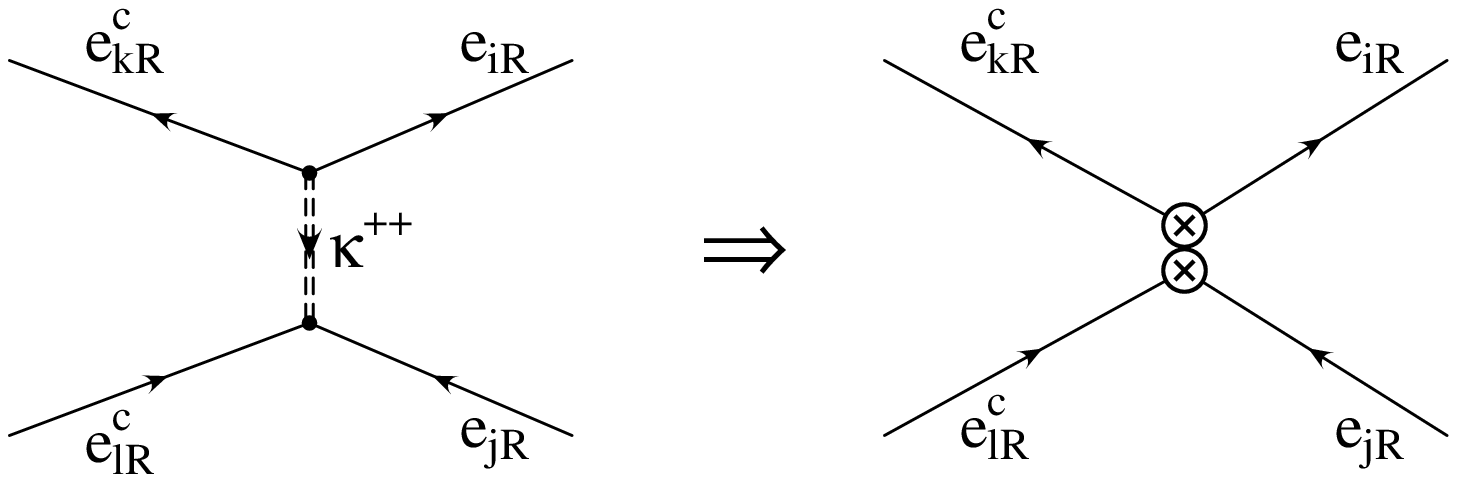}{Tree-level matching.}{fig:tree-match}
This interaction is of the type $\lrr$.
Comparing with \eq{eq:rr} one can immediately identify
\begin{equation}
\alpha_{ik;lj}^{RR} = h^*_{ik} h_{lj}\bla {\mathrm{and}}
\bla \Lambda = m_\kappa\,.
\label{eq:alphak}
\end{equation}

On the other hand, by using the techniques in \cite{misha} one can easily
obtain the contributions from matching to the full theory at one
loop to the rest of the $\alpha$'s. The one loop diagrams involving
$\kappa^{++}$ are depicted in fig.~\ref{fig:loop-match}.
\mafigura{12cm}{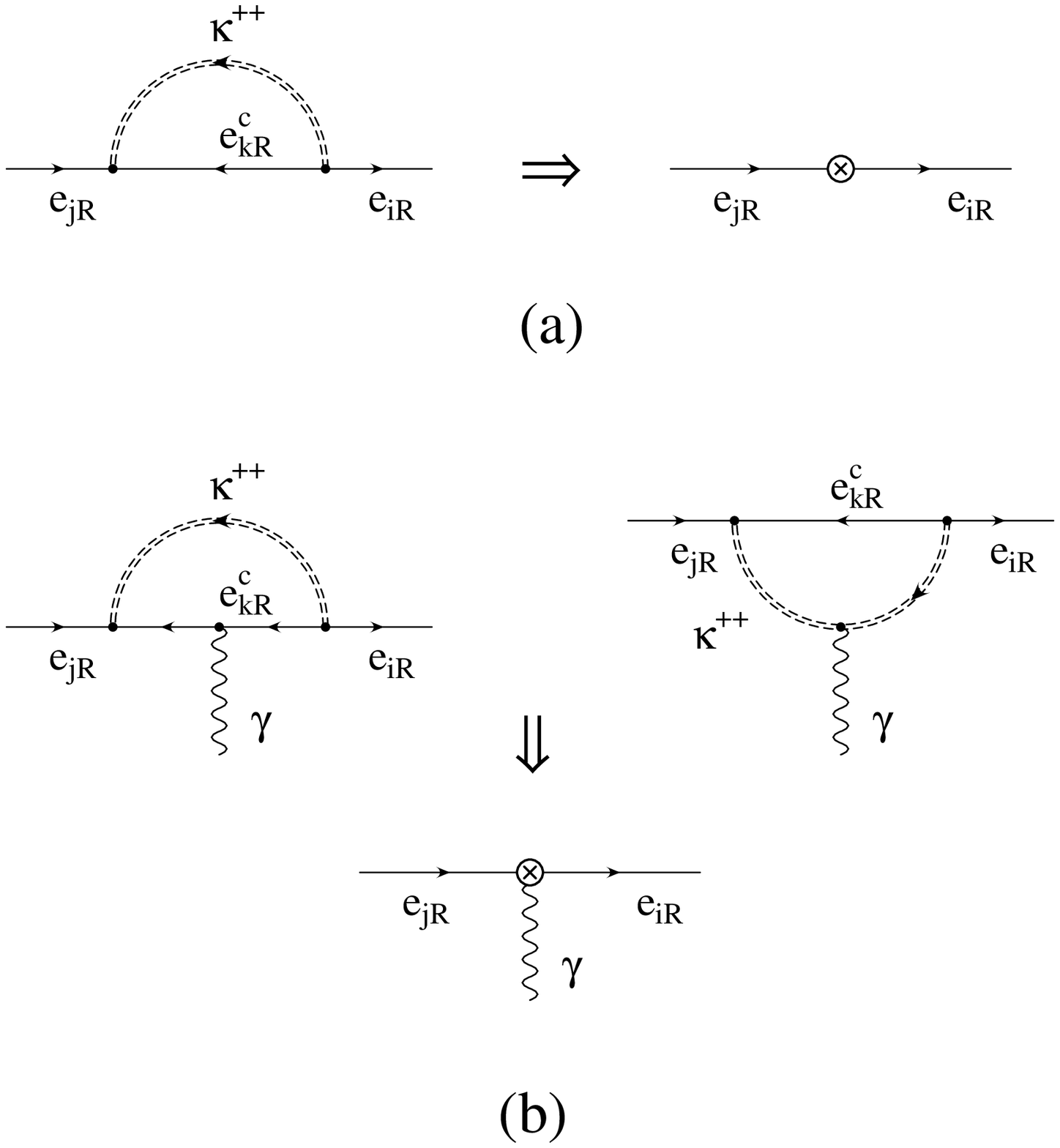}{One-loop matching.}{fig:loop-match}
Those diagrams are computed by using dimensional regularization
and, after subtraction of the effective theory contributions,
fig. \ref{fig:effective-loop}, they can be expanded in $1/m^2_\kappa$. We keep
at most terms of order $1/m^2_\kappa$. At this order 
there are contributions to the self-energies and to the vertex of
the photon. Those give three types of operators:
charge radius operators, \eq{eq:la} and \eq{eq:ra}, magnetic moment
operators, \eq{eq:sigmal} and \eq{eq:sigmar}, and operators that 
involve three covariant derivatives of the fermions. The last operators can
be removed by using the equations of motion in favour of mass terms
and do not lead to any interesting physics. Therefore, after wave function
renormalization, in order to write the kinetic terms in canonical form, the 
only operators generated in this model are those appearing in the lagrangian 
(\ref{eq:leff}), but only the right-handed components. 
By using the $\overline{\mathrm{MS}}$ renormalization scheme
and by choosing the renormalization scale $\mu=\Lambda=m_\kappa$, we 
obtain the following coefficients 
\begin{eqnarray}
\alpha_{ij}^{R} & = & 
\frac{20 \,h^*_{ki} h_{kj}}{9 }, 
\label{asfull} \\
\alpha_{ij}^{\sigma R} & = & \frac{2 \,h^*_{ki} h_{kl}}{3}\, . 
\label{eq:asigma}
\end{eqnarray}
Adding up the contributions from these operators and the contributions
coming from the diagram in fig.~\ref{fig:effective-loop},
i.e., substituting eqs. \rfn{asfull}, \rfn{eq:asigma} in
eqs.~\rfn{eq:fv}--\rfn{eq:sigmaa}, we obtain exactly the
same amplitudes as obtained from a calculation in the full model
up to terms ${\cal O}((m_\mu/m_\kappa)^4)$. Therefore  the full
amplitude for \mec{} is dominated by the divergent contribution from
the diagram in fig.~\ref{fig:effective-loop}. In the effective field theory
language one says that the amplitude is dominated by the running from the
scale of new physics $\Lambda= m_\kappa$ to relevant scale of the process:
in the case of \mec{}, $m_\mu$ (or $m_\tau$ if $\tau$ leptons are
running in the loop). This conclusion is independent on the model as 
long as four-fermion  interactions \eq{eq:ll} and/or \eq{eq:rr} exist.

As expected, in this model there are no contributions to
$\overline{e_R} \gamma_\mu e_R Z^\mu$ since
the physics of a scalar singlet decouples for $m_\kappa \gg \Lambda_F$. 
This means that the contributions of the scalar can only lead to
operators  suppressed by, at least, $1/m_\kappa^2$. For instance
one could easily obtain an operator like \eq{eq:ra} with the
photon replaced by the $Z$-boson. The contribution of those operators
to \mec{}, however, are suppressed by a factor $(m_\mu/m_Z)^2$ with 
respect to the photonic contributions.

\section{Numerical results and conclusions}

Now we are ready to compare the branching ratios of \mec{} and \meg{}
in the class of models we consider and to analyze the relative
potential of different experiments to test muon flavour conservation.
For definiteness we consider only the right-handed operators in 
lagrangian \rfn{eq:leff}.
We first study the general case in the
framework of our effective theory and then we
apply the results to our specific model.
Substituting  the form factors \rfn{eq:fv}-\rfn{eq:sigmaa}
to \eq{eq:xi} and taking only the dominant logarithmic terms
(with  $\Lambda=1$ TeV in the logarithms)  we find  
\bea
R^{Ti}_{\mu e}& =& 1.2\;(0.5)
\cdot 10^{-5}\;\mrm{TeV}^{4}\;
\left(\frac{\alpha_{ke;k\mu}^{RR}}{\Lambda^2}\right)^2
\,,
\label{tirate} \\
R^{Pb,\,Au }_{\mu e}& = & 3.5\;(1.4)
\cdot 10^{-5}\;\mrm{TeV}^{4}\;
\left(\frac{\alpha_{ke;k\mu}^{RR}}{\Lambda^2}\right)^2
\,,
\label{pbrate}
\eea
where the first number in the expressions corresponds to $k=e,\;\mu$ and
the number in the brackets to $k=\tau$.
If $k=\tau$ the logarithm enhancement is $\ln(m_\tau/\Lambda)$
instead of $\ln(m_\mu/\Lambda)$ and it is slightly smaller than
in the $\mu$ or $e$ cases.
Note that the conversion is somewhat enhanced in $Pb$ and
$Au$ (in fact, maximized \cite{chiang})  if compared with $Ti$.
 
In the effective lagrangian framework \meg{} does not get contributions 
from loops in fig. \ref{fig:effective-loop} and, therefore, it is not enhanced
by large logarithms. In any full theory in 
which both \mec{} and \meg{}
are induced by loops all the couplings  
should be of the same magnitude (compare, e.g., 
\eq{eq:alphak} with \eq{eq:asigma} in our doubly charged scalar model).  
Assuming 
$\alpha^{\sigma R}\equiv\alpha^{RR}$  we obtain
\begin{equation}
R_{\gamma} = 1.2\cdot 10^{-5} \;\mrm{TeV}^{4}\;
\left(\frac{\alpha_{ke;k\mu}^{RR}}{\Lambda^2}\right)^2
\,
\label{megrate2}
\end{equation}
for any type of fermion in the loop. 

Comparison of eqs. \rfn{tirate}, \rfn{pbrate} 
with \eq{megrate2} shows that due to the presence of 
large logarithms the \mec{} rate 
is comparable or even exceeds the \meg{} rate. 
To constrain new physics we have to also 
take into account the sensitivity of experiments.
The present experimental upper 
limits on the branching ratios of the processes are
$R^{Ti}_{\mu e}(exp)\lsim 4.3\cdot 10^{-12}$ \cite{psiti},
$R^{Pb}_{\mu e}(exp)\lsim 4.6\cdot 10^{-11} $ \cite{psipb} and
$R_{\gamma}(exp)\lsim 4.9\cdot 10^{-11}$ \cite{pdb}.
SINDRUM II experiment at PSI taking presently data  on gold will reach the 
sensitivity  $R^{Au,\, expected}_{\mu e}\lsim 5\cdot 10^{-13}$ \cite{imprau}
and starting next year the final run on $Ti$ it should reach
 $R^{Ti,\, expected}_{\mu e}\lsim 3\cdot 10^{-14}$ \cite{imprti}.
Normalizing the branching ratios to the experimental upper limits
we get for $\Lambda=1$ TeV
\bea
\frac{R^{Ti}_{\mu e} }{4.3\cdot 10^{-12}}&=&11.4\;(5.7)\;B^{Ti}\,
\frac{R_{\gamma}}{4.9\cdot 10^{-11}} \,, 
\label{ratioti}\\
\frac{R^{Pb,\, Au}_{\mu e} }{4.6\cdot 10^{-11}}&=&3.1\;(1.25)\;B^{Au}\,
\frac{R_{\gamma}}{4.9\cdot 10^{-11}} \,, 
\label{ratiopb}
\eea
where, again, numbers in the brackets correspond to the case $k=\tau$
and the factors $B,$ $B=R_{\mu e}^{present}(exp)/R_{\mu e}^{future}(exp),$ 
take into 
account the improvements in the experimental sensitivity. 
\Eq{ratioti} and \eq{ratiopb} constitute the central result of this work:
in the class of models we consider searches for \mec{} in both $Ti$ and $Pb$
constrain new physics more 
stringently than searches for \meg{}. 
In addition, from SINDRUM II one expects to achieve $B^{Au}=92.$ 
already  in forthcoming months and $B^{Ti}=1.4\cdot 10^{2}$ next year. 
If the aimed sensitivity will be achieved then $\mu$--$e$ experiments 
probe the couplings $\alpha$  of new physics more than one 
order of magnitude more stringently than $\mu\rightarrow e\gamma.$

\begin{table}
\begin{tabular}{|c|c|c|c|c|}
\hline\hline
$R=$ & $4.6\cdot 10^{-11}$ &$4.3\cdot 10^{-12}$&$5.0\cdot 10^{-13}$&
$3.0\cdot 10^{-14}$ \\
\hline\hline
log-enhanced $\mu$--$e$ & 
32 & 44 &101 & 158
\\
\hline
non-enhanced $\mu$--$e$ &
7 & 9 & 20 & 32 \\
\hline
$\mu\rightarrow e\gamma$ & 23 & 41 & 70 & 141
\\
\hline\hline
\end{tabular}
\caption{Values of  $\Lambda$, in TeV, probed in \mec{} and \meg{} for different
upper bounds on the branching ratios. The upperbound $4.6\cdot 10^{-11}$
is the present bound for \mec{} on $Pb$ and it is very close to the present
\meg{} bound ($4.9\cdot 10^{-11}$). $4.3\cdot 10^{-12}$ is the present
bound for \mec{} on $Ti$. $5\cdot 10^{-13}$ and $3\cdot 10^{-14}$ are
the expected bounds in the next year for \mec{} on $Au$ and $Ti$,
respectively.
}
\end{table}

To show which scales of new physics $\Lambda$ can be probed in
\mec{} and \meg{} experiments we have presented the values of $\Lambda$
in TeV-s in Table 1 for different experimental upper bounds
on the branching ratios  of the processes. 
All the couplings $\alpha$ are  taken to 
be equal to unity. We have considered both classes of models
with and without logarithmic enhancement of \mec{}.   
If the experimental limits for \mec{} and \meg{} are equal then 
\mec{} enhanced by large logarithms has 
better sensitivity to $\Lambda$ than \meg{}, especially 
in the case of $Pb$ and $Au$ experiments. 
The scales testable reach $\Lambda\sim {\cal O}(10^2)$ TeV.
However, \mec{} without logarithmic enhancement can only probe 
scales lower by about a factor 5.

To illustrate the discussion above let us present the experimental
bounds on the couplings $h$ of $\kappa^{++}$ in our model. 
Substituting  the couplings in 
eqs. \rfn{eq:alphak}-\rfn{eq:asigma} to
eqs. \rfn{eq:fv}-\rfn{eq:sigmaa}
 and using the present experimental limit for $Ti$
we obtain from \mec{} for $m_\kappa=1$ TeV
\bea
h_{e\mu}h_{ee}^*\;,\;h_{\mu\mu}h_{e\mu}^*\;&\lsim&\;
\frac{6\cdot 10^{-4}}{\sqrt{B^{Ti}}}\,, 
\nn \\
h_{\tau\mu}h_{e\tau}^*\;&\lsim&\;\frac{9\cdot 10^{-4}}{\sqrt{B^{Ti}}}\,, 
\label{hk++}
\eea
while $\mu\rightarrow e\gamma$ gives
\bea
h_{k\mu}h_{ek}^*\;&\lsim&\;3\cdot 10^{-3}\, .
\eea
The bounds \rfn{hk++} are new limits
on the off-diagonal doubly charged scalar interactions (note that 
tree level $\mu\rightarrow 3e$ probes only $h_{\mu e}h_{ee}^*$). 
While derived for
the right-handed singlet the limits apply with a good accuracy also for
the interactions of triplet scalars appearing in models with enlarged
Higgs sectors  as well as in left-right symmetric 
models
This is because the doubly charged component
of triplet gives the  dominant contribution both to \mec{} and \meg{}.
Note that the upper bounds \rfn{hk++} 
are going to be improved by an order of magnitude with new \mec{} data.

Finally,  we would like to stress that our main result, the logarithmic
enhancement of \mec{} rate, is completely general and
applies to all models with effective  interactions of four 
charged fermions. For simplicity we have constrained ourselves  to 
purely leptonic operators. However, the same effect is also present 
for operators involving quarks. To get large logarithms one just needs light 
charged fermions in the loop. Therefore, loop induced 
\mec{} is also enhanced in models with broken $R$-parity \cite{muerparity} and 
leptoquarks but not in $R$-conserving MSSM or SUSY GUT's considered in 
ref.~\cite{vergados} in which the light fermions in loops are necessarily 
neutral.


In conclusion, using the effective lagrangian description of new physics we have
pointed out a wide class of models with effective four charged fermion
interactions  in which loop induced $\mu$--$e$ conversion in nuclei is 
enhanced by large logarithms. With the present upper limits on  
$\mu$--$e$ conversion and $\mu\rightarrow e\gamma$ branching ratios
bounds on new physics (occurring at loop level) derived from these processes
are more restrictive in the case of  $\mu$--$e$ conversion. 
In nearest future  this factor will increase by more than one
order of magnitude due to the expected improvements in sensitivity  of 
already running $\mu$--$e$ conversion experiments. This
general result is confirmed by  exact calculations in the extension of the 
SM with doubly charged singlet scalar.

\begin{ack}
We thank A. van der Schaaf for information about ongoing experiments at PSI
and H.C. Chiang and E. Oset for discussions on nuclear physics involved
in $\mu$--$e$ conversion.
A.S. thanks SISSA, Trieste (Italy) for the warm hospitality during
his visit there, where part of this work has been done.
This work has been supported in
part by CICYT (Spain) under the grant AEN-96-1718.
\end{ack}


\begin{thebibliography}{99}

\bibitem{psiti} SINDRUM II Collaboration, C. Dohmen \ea, 
\pl{B317} (1993) 631.

\bibitem{psipb}  SINDRUM II Collaboration, W. Honecker \ea, 
\prl{76} (1996) 200.

\bibitem{triumph} S. Ahmad \ea, \pr{D38} (1988) 2102.

\bibitem{imprti} A. van der Schaaf (spokesman of SINDRUM II), PSI proposal
R-87-03, 1987.

\bibitem{proposal} M. Bachman \ea, BNL Proposal P940, 1997;
for a review see also, A. Czarnecki, hep-ph/9710425.

\bibitem{reviews} For original references see, e.g.,
J.D. Vergados, \prep{133} (1986) 1;
S.M. Bilenky and S.T. Petcov, \rmp{59} (1987) 671.  

\bibitem{vergados} See, e.g., 
R. Barbieri and L. Hall, \pl{B338} (1994) 212; 
R. Barbieri, L. Hall and A. Strumia, \np{B445} (1995) 219;
T.S. Kosmas and J.D. Vergados, \prep{264} (1996) 251, and references 
therein.

\bibitem{altarelli} W.J. Marciano and A.I. Sanda, \prl{38} (1977) 1512;
G. Altarelli \ea, \np{B125} (1977) 285.

\bibitem{triplets}
  T.P. Cheng, L. Li, 
   Phys.~Rev. {\bf D22} (1980) 2860;
  G.B. Gelmini, M. Roncadelli, 
   Phys.~Lett. {\bf B99} (1981) 411;
   H.M.~Georgi, S.~Glashow and S.~Nussinov, \np{B193} (1981) 297.

\bibitem{lr}
  J.C. Pati, A. Salam,  Phys.~Rev. {\bf D10} (1974) 275; 
  R.N. Mohapatra, J.C. Pati, 
   Phys. Rev. {\bf D11} (1975) 566, {\it ibid.} 2558; 
  G. Senjanovic, R.N. Mohapatra,  Phys. Rev. {\bf D12} (1975) 1502; 
 R.N. Mohapatra, G. Senjanovic, \prl{44} (1980) 912;
 R.N. Mohapatra, G. Senjanovic,  Phys. Rev. {\bf D23} (1981) 165.


\bibitem{rparity} C.S. Aulakh and R.N. Mohapatra, \pl{B119}(1982) 136;
L.J. Hall and M. Suzuki, \np{B231} (1984) 419; 
J. Ellis \ea, \pl{B150} (1985) 142;
G.G. Ross and J.W.F. Valle, \pl{B151} (1985) 375;
S. Dawson, \np{B261} (1985) 297;
R. Barbieri and A. Masiero, \np{B267} (1986) 679. 

\bibitem{leptoquarks} W. Buchm\"uller, R. R\"uckl and D. Wyler, 
\pl{B191} (1987) 442.

\bibitem{megtriplet} J. Bernab\'eu, A. Pich and A. Santamaria,
\pl{148B} (1984) 229.

\bibitem{zee} A. Zee, \pl{93B} (1980) 389;
S.T. Petcov, \pl{115B} (1982) 401.

\bibitem{misha} M. Bilenky and A. Santamaria, \np{B420} (1994) 47.

\bibitem{doublycharged} K. Babu, \pl{B203} (1988) 132.

\bibitem{eqmotion} See for instance, H. Georgi, \np{B361} (1991) 339;
H. Simma, \zp{C61} (1994) 67; C. Arzt, \pl{B342} (1995) 189.

\bibitem{ng1}
D. Ng and J.N. Ng, \pl{B320} (1994) 181; \pl{B331} (1994) 371.

\bibitem{zbbar} J. Bernab\'eu, A. Pich and A. Santamaria, \pl{200B} (1988)
569; \np{B363} (1991) 326. For an effective lagrangian calculation of
radiate corrections to $Z\rightarrow b\bar{b}$ see S. Peris and A. Santamaria, 
\np{B445} (1995) 252.

\bibitem{nondec}
D. Ng and J.N. Ng, \pl{B331} (1994) 371; 
D. Tommasini \ea, \np{B444} (1995) 451; 
G. Barenboim and M. Raidal, \np{B484} (1997) 63.

\bibitem{pont} S.M. Bilenky, S.T. Petcov and B. Pontecorvo,
\pl{67B} (1977) 309.

\bibitem{zprim}
J. Bernabeu, E. Nardi and D. Tommasini, \np{B409} (1993) 69.

\bibitem{wf} S. Weinberg and G. Feinberg, \prl{3} (1959) 111.

\bibitem{ver1} T.S. Kosmas and J.D. Vergados, \np{A510} (1990) 641.

\bibitem{chiang} H.C. Chiang \ea, \np{A559} (1993) 526.

\bibitem{ver2} T.S. Kosmas, Amand Faessler and J.D. Vergados, 
nucl-th/9704020. 

\bibitem{kim}
 J.E. Kim, P. Ko and D.-G. Lee, \pr{D56} (1997) 100.

\bibitem{davidson}
S. Davidson, D. Bailey and B. Campell, \zp{C61} (1994) 613. 

\bibitem{chiang2} H.C. Chiang, E. Oset and P. Fernandez de Cordoba,
\np{A510} (1990) 591.

\bibitem{imprau} A. van der Schaaf (spokesman of SINDRUM II), private 
communication.


\bibitem{pdb} R.M. Barnett \ea, {\it Review of Particle Physics,}
\pr{D54} (1996) 1.

\bibitem{muerparity} K. Huitu, J. Maalampi, M. Raidal and A. Santamaria, 
FTUV/97-45, to appear elsewhere.

\end{thebibliography}
\end{document}